\documentclass[pdflatex,sn-mathphys-num]{sn-jnl}

\usepackage{graphicx}%
\usepackage{multirow}%
\usepackage{amsmath,amssymb,amsfonts}%
\usepackage{amsthm}%
\usepackage{mathrsfs}%
\usepackage[title]{appendix}%
\usepackage{xcolor}%
\usepackage{textcomp}%
\usepackage{manyfoot}%
\usepackage{booktabs}%
\usepackage{algorithm}%
\usepackage{algorithmicx}%
\usepackage{algpseudocode}%
\usepackage{listings}%

\theoremstyle{thmstyleone}%
\theoremstyle{thmstyletwo}%

\theoremstyle{thmstylethree}%

\begin{document}

\title[Article Title]{Ultra-Low-Noise Brillouin Hybrid Synthetic Laser for Sub-Hertz Lattice Clock Spectroscopy}


\author[1]{\fnm{Meiting} \sur{Song}}\email{msong109@ucsb.edu}
\author[2,3]{\fnm{Stefan} \sur{Lannig}}\email{stefan.lannig@jila.colorado.edu}
\author[2,3]{\fnm{Dahyeon} \sur{Lee}}\email{dahyeon.lee@colorado.edu}
\author[2,3]{\fnm{Lingfeng} \sur{Yan}}\email{lingfeng.yan@colorado.edu}
\author[1]{\fnm{Andrei} \sur{Isichenko}}\email{aisichenko@ucsb.edu}
\author[1]{\fnm{Nick} \sur{Montifiore}}\email{nmontifiore@ucsb.edu}
\author[3,4]{\fnm{Nitesh} \sur{Chauhan}}\email{nitesh.chauhan@colorado.edu}
\author[2,3]{\fnm{Max N.} \sur{Frankel}}\email{max.frankel@colorado.edu}
\author[2,3]{\fnm{Yu Hyun} \sur{Lee}}\email{yuhyun.lee@colorado.edu}
\author[2,3]{\fnm{Shraddha} \sur{Agrawal}}\email{shraddha.agrawal@colorado.edu}
\author*[2,3]{\fnm{Jun} \sur{Ye}}\email{ye@jila.colorado.edu }
\author*[1]{\fnm{Daniel J.} \sur{Blumenthal}}\email{danb@ucsb.edu}

\affil[1]{\orgdiv{Department of Electrical and Computer Engineering}, \orgname{University of California Santa Barbara}, \orgaddress{\city{Santa Barbara}, \state{California} \postcode{93106}, \country{USA}}}

\affil[2]{\orgdiv{JILA}, \orgname{National Institute of Standards and Technology and University of Colorado}, \orgaddress{\city{Boulder}, \state{Colorado} \postcode{80309}, \country{USA}}}

\affil[3]{\orgdiv{Department of Physics}, \orgname{University of Colorado}, \orgaddress{\city{Boulder}, \state{Colorado} \postcode{80309}, \country{USA}}}

\affil[4]{\orgdiv{Time and Frequency Division}, \orgname{National Institute of Standards and Technology}, \orgaddress{\city{Boulder}, \state{Colorado} \postcode{80309}, \country{USA}}}


\abstract{Frequency-stable lasers enable high-fidelity quantum state manipulation, which forms the basis of optical atomic clocks, quantum sensing, and quantum computation. Performing state manipulations at increasingly high speeds requires attention to laser frequency noise at high Fourier (carrier-offset) frequencies that cannot be addressed by traditional cavity stabilization alone. Scalable operations also benefit from device miniaturization. Here, we demonstrate a hybrid laser stabilization approach that combines ultrahigh frequency stability of a cryogenic silicon cavity with high-Fourier-frequency noise suppression of an integrated Brillouin laser. The combined system suppresses frequency noise over a Fourier span of more than 7 decades, yielding a $<$1 Hz phase-integrated linewidth and 0.2 Hz$^2$/Hz frequency noise at Fourier frequencies above 10 MHz. The performance of this hybrid laser is confirmed by sub-Hz Rabi spectroscopy with a three-dimensional $^\text{87}$Sr lattice clock. This work demonstrates record-low frequency noise at 698 nm over an extensive Fourier frequency range and highlights the promise of precision clock spectroscopy using a chip-scale integrated laser technology.}

\maketitle

\section{Introduction}\label{sec1}

Frequency-stabilized lasers based on high-finesse cavities are a foundational component in optical atomic clocks \cite{ludlow_optical_2015, marshall_highstability_2025, yang_clock_2025}, quantum information processing \cite{ladd_quantum_2010, degen_quantum_2017, bruzewicz_trappedion_2019}, and gravitational wave detection \cite{aasi_advanced_2015, kwee_stabilized_2012, meylahn_stabilized_2022}. Continued advances in cavity engineering over the last few decades have firmly established cavity-stabilized lasers as the gold standard for long term spectral purity, with cryogenic silicon (Si) cavities representing the state-of-the-art \cite{lee_frequency_2026}. However, conventional cavity stabilization cannot effectively suppress laser frequency noise above 10 MHz due to the finite response of frequency modulators, with feedback bandwidth limited to $<$1 MHz and feedforward extending suppression to 10 MHz \cite{bagheri_semiconductor_2009, li_active_2022, chao_pound_2024, denecker_measurement_2025}. Such high Fourier frequency noise has become increasingly important as quantum computation and sensing speeds and the rate of qubit manipulations are increased.  

In parallel with the development of cavity-stabilized lasers, several photonic integrated stabilized lasers, most notably integrated silicon nitride (Si$_3$N$_4$) Brillouin lasers \cite{liu_ultralow_2022,puckett_422_2021}, have shown frequency noise suppression at high Fourier frequencies well beyond the reach of electronic feedback. This approach achieves broadband suppression of fast pump laser frequency noise (typically $>$10 kHz) with stimulated Brillouin scattering (SBS) \cite{debut_linewidth_2000, li_characterization_2012, behunin_fundamental_2018, yuan_linewidth_2020, loh_ultralow_2024} through the wide Brillouin gain bandwidth, long effective Brillouin grating lengths, and highly nonlinear parametric gain \cite{liu_large_2025, shepherd_noise_2026}. While the Brillouin process is effective at suppressing the high-Fourier-frequency noise, low- to mid-Fourier frequency noise is limited by thermorefractive noise (TRN) and other sources such as photothermal noise in the laser cavity. Therefore, hybrid laser strategies are needed to mitigate noise over large low- to high-Fourier-frequency ranges \cite{liu_photonic_2022, guo_chipbased_2022, song_octave_2026}.

\begin{figure*}[h!]
\centering
\includegraphics[width=1.0\textwidth]{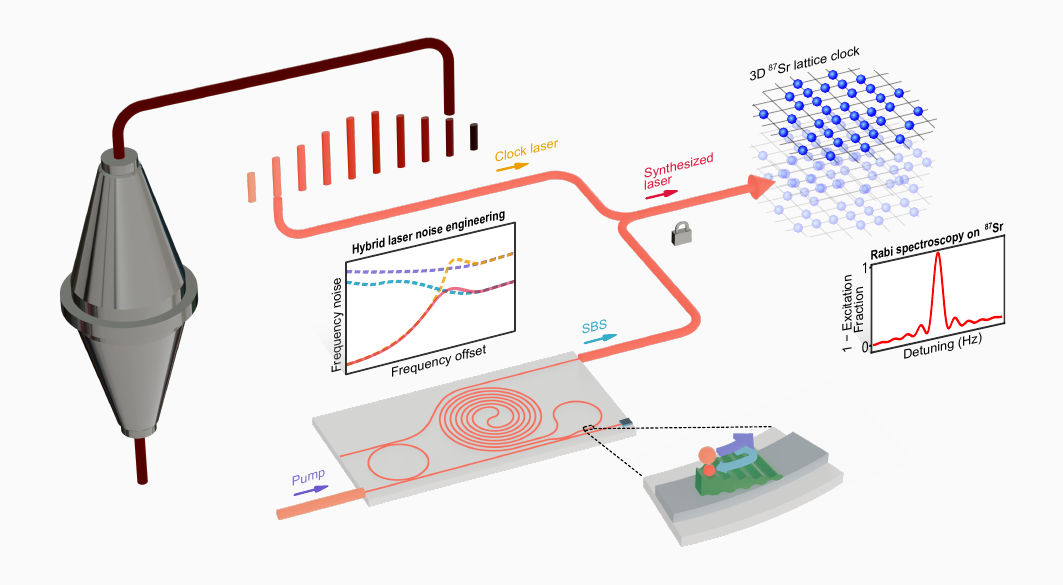}
\caption{Illustration of the hybrid synthetic laser. A 1542 nm laser is stabilized to the cryogenic Si cavity and the stability is transferred to 698 nm through a frequency comb. An integrated Si$_3$N$_4$ photonic chip pumped at 698 nm generates the Brillouin laser. Two schematic plots illustrate the hybrid laser frequency noise, highlighting noise suppression at different stages (see Fig.~\ref{fig4}), and Rabi spectroscopy of $^{87}$Sr atoms trapped in a three-dimensional lattice (see Fig.~\ref{fig5}). The zoomed-in waveguide section depicts the Brillouin scattering process, in which a pump photon (light red sphere and purple arrow) generates an acoustic phonon (green) and a back-scattered Stokes photon (darker red sphere and blue arrow). The SBS laser is phase-locked to the Si-cavity-stabilized laser using an optical phase-locked loop, and the synthetic laser output is delivered to a $^{87}$Sr three-dimensional optical lattice clock for Rabi spectroscopy measurements. Components are not drawn to scale. }\label{fig1}
\end{figure*}

This frequency noise landscape motivates a hybrid strategy that combines SBS-based high-Fourier-frequency noise suppression with precision cavity stabilization. Here, we report a hybrid synthetic laser that combines a photonic-chip-based, self-isolating 65-cm-long 698 nm Brillouin resonator with a state-of-the-art cryogenic Si cavity, as schematically depicted in Fig.~\ref{fig1}. This design enables engineering of the frequency noise over a Fourier span of more than 7 decades, simultaneously achieving a sub-Hz phase-integrated linewidth and a frequency noise power spectral density of 0.2 Hz$^2$/Hz at Fourier frequencies above 10 MHz. The resulting linewidth and frequency noise are validated using in-loop error signal analysis and sub-Hz Rabi spectroscopy on the clock transition in a three-dimensional $^{87}$Sr optical lattice clock. This level of multi-decade broadband frequency noise suppression and enhanced stability unlocks the potential of low-noise lasers to advance high-speed gate operations for quantum computing and sensing.

\section{Components, Integration, and System Characterization}\label{sec2}

\begin{figure*}[h!]
\centering
\includegraphics[width=1.0\textwidth]{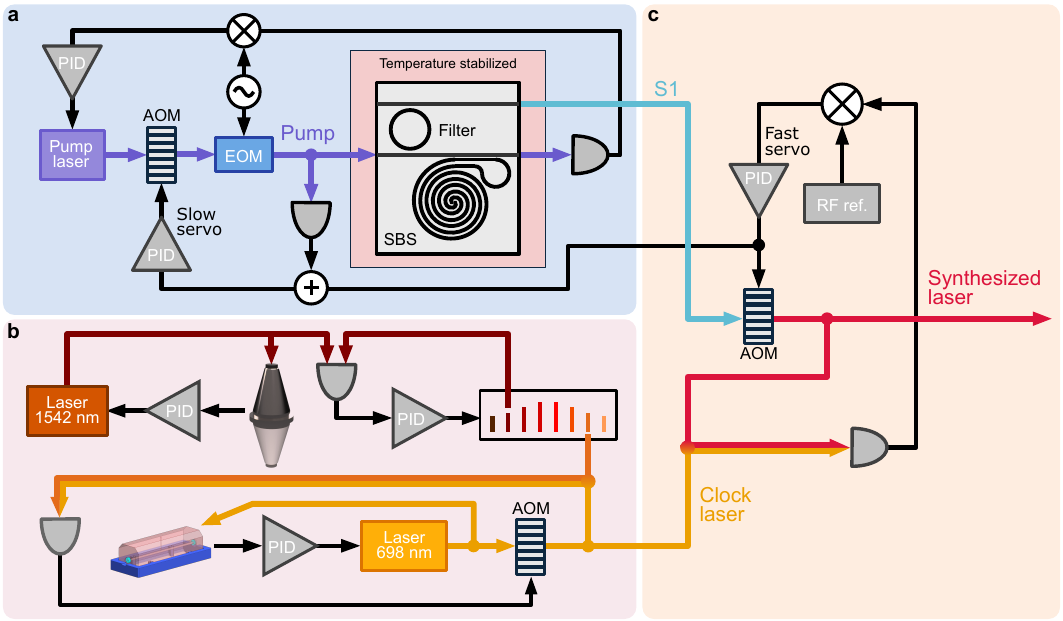}
\caption{\textbf{System schematic of the hybrid synthetic laser.} \textbf{a} The pump laser (purple) is locked to the on-chip Brillouin cavity using a Pound–Drever–Hall loop. An integrated filter ring routes the backscattered Brillouin signal (blue) to the drop port. \textbf{b} The clock laser (orange) is a ULE-cavity-stabilized laser locked to a Si-cavity referenced laser through an optical comb. \textbf{c} The Brillouin laser is subsequently stabilized to the clock laser using an optical phase-locked loop to form the hybrid synthetic laser (red).}\label{fig2}
\end{figure*}

The hybrid synthetic laser combines a cryogenic Si reference cavity with an integrated Brillouin laser based on a low-loss, high-Q Si$_3$N$_4$ resonator. The system schematic in Fig.~\ref{fig2} depicts the two-stage laser stabilization. For low- to mid-Fourier-frequency noise suppression and stabilization synthesis, a 1542 nm laser is referenced to the cryogenic Si cavity. Its stability is transferred via an optical frequency comb to a 698 nm laser referenced to an ultra-low-expansion (ULE) cavity \cite{yan_high-power_2025}, hereafter referred to as the “clock laser”. The Brillouin laser that suppresses high-Fourier-frequency noise is subsequently phase-locked to the clock laser, forming the hybrid synthetic laser that inherits the low-noise performance of both subsystems. 

The high-Fourier-frequency noise suppression in the Brillouin laser arises from phonon-mediated interactions between long-lived intra-cavity pump photons and rapidly decaying acoustic modes, enhanced by the large nonlinear interaction length of the ultra-high-Q coil resonator. This mechanism attenuates pump noise at bandwidths well beyond those accessible to conventional electronic feedback. In contrast to other integrated Brillouin laser designs that utilize phase matched phonon and photon guiding, the unguided free-space-like rapidly decaying acoustic waves provide a large Brillouin gain bandwidth (hundreds of MHz) and remove the propagation delay phase matching requirement between acoustic and optical modes \cite{gundavarapu_subhertz_2019} enabling integrated Brillouin lasers from the visible to the shortwave infrared. 

\begin{figure*}[h!]
\centering
\includegraphics[width=1.0\textwidth]{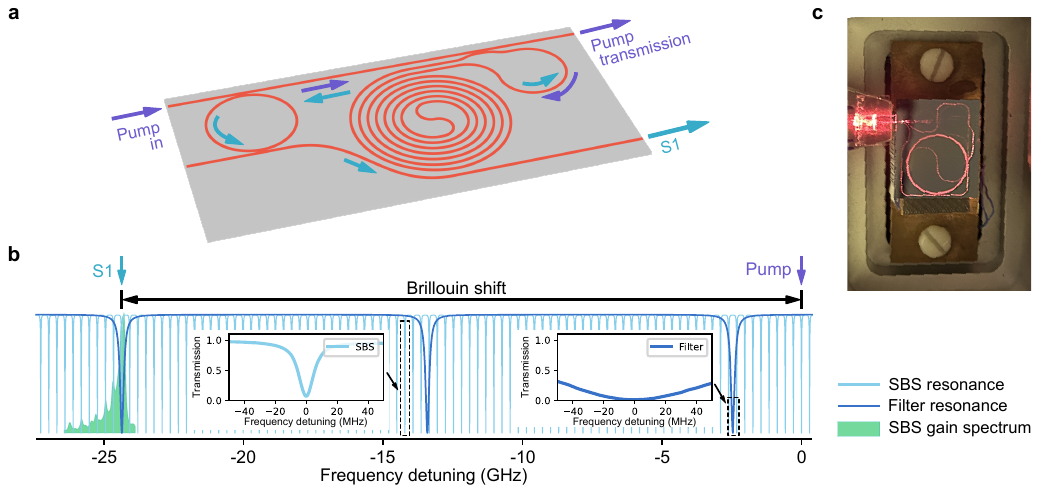}
\caption{\textbf{Integrated photonic device combining SBS and filter resonators for large mode volume and circulator-free operation.} \textbf{a} Schematic of the integrated photonic circuit showing pump and S1 propagation. The pump traverses the filter resonator off resonance and excites S1 via SBS in a 65-cm-long resonator. The generated S1 is then spectrally filtered and routed by the filter resonator. Arrows indicate propagation direction.  \textbf{b} Calculated transmission spectra of the device over a wide frequency detuning range, based on single-resonance and FSR measurements. The shaded region denotes the SBS gain spectrum, and the frequency separation between the pump and the selected signal corresponds to the Brillouin shift. The FSR of the narrow SBS resonances (light blue) is an integer divisor of the Brillouin shift, whereas the filter resonance (dark blue) is intentionally incommensurate with it. Insets: measured transmission spectra of the SBS resonance (left) and the filter resonance (right) versus frequency detuning. \textbf{c} Photograph of the device under test, edge-coupled using a fiber array. The bright coil resonator results from scattering of resonantly enhanced light in the Brillouin resonator. The filter ring in the upper right of the chip is less visible because it is off resonance with the pump wavelength. }\label{fig3}
\end{figure*}

 The Si$_3$N$_4$ Brillouin coil resonator has a cavity length of 65 cm and is co-integrated with an first-order Stokes (S1) add-drop filter that provides Stokes output separation from the pump (Fig.~\ref{fig3}a). The 698-nm pump laser is locked to the integrated resonator (see Fig.~\ref{fig2}a and Methods for details). This design yields an on-chip lasing threshold of 58 mW, 32 mW Brillouin laser power at 154 mW pump power, and high-Fourier-frequency noise down to 0.2 Hz$^2$/Hz. The long cavity lowers the high-Fourier-frequency noise floor and raises the threshold for higher-order Stokes generation, allowing S1 power to increase monotonically with pump power without second-order Stokes (S2) generation or frequency-noise back-action \cite{behunin_fundamental_2018}. The choice of cavity length is a balance among the Brillouin lasing threshold, output power, and resulting high-Fourier-frequency noise suppression. The add-drop filter ring is resonant at S1 and non-resonant at the pump wavelength, routing the backscattered S1 light to the drop port. As seen in Fig.~\ref{fig3}b, the filter ring, with an 11 GHz FSR and a 154 MHz resonance linewidth, is sufficiently spaced to avoid overlap with the 24.4 GHz Brillouin shift. This design provides over 25 dB of self-isolation \cite{liu_circulatorfree_2023} for the integrated Brillouin laser, serving as a built-in circulator function and enables future integration with other Si$_3$N$_4$ components. Additionally, this level of integration reduces losses, minimizes glass-air interfaces, and lowers the risk of optical surface damage at high pump powers. The photonic chip (Fig.~\ref{fig3}c) is fabricated using the ultra-low loss thin core process \cite{blumenthal_silicon_2018} with details of design and fabrication provided in Methods.

The low-Fourier-frequency-stable reference (clock) laser is referenced to a cryogenic Si cavity and a room-temperature ULE cavity \cite{yan_high-power_2025} (see Fig.~\ref{fig2}b). The stability of the SBS laser and the Si cavity is combined with an optical phase-locked loop (OPLL) shown in Fig.~\ref{fig2}c. The OPLL feedback signal actuates an acousto-optic modulator (AOM), which cancels the low-Fourier-frequency thermal fluctuations of the Brillouin laser and thus transfers the low-Fourier-frequency clock laser stability to the S1 light. At Fourier frequencies beyond the OPLL bandwidth the noise performance of the Brillouin laser is preserved. Slow thermal drifts of the Brillouin resonator at the mK level can saturate the OPLL dynamic range of $\sim10$ MHz. Therefore, the OPLL includes an additional slow feedback with large dynamic range onto the pump laser intensity, tuning the SBS resonator by effectively acting as an intra-cavity thermal actuator.  

\begin{figure}[h!]
\centering
\includegraphics[width=1.0\textwidth]{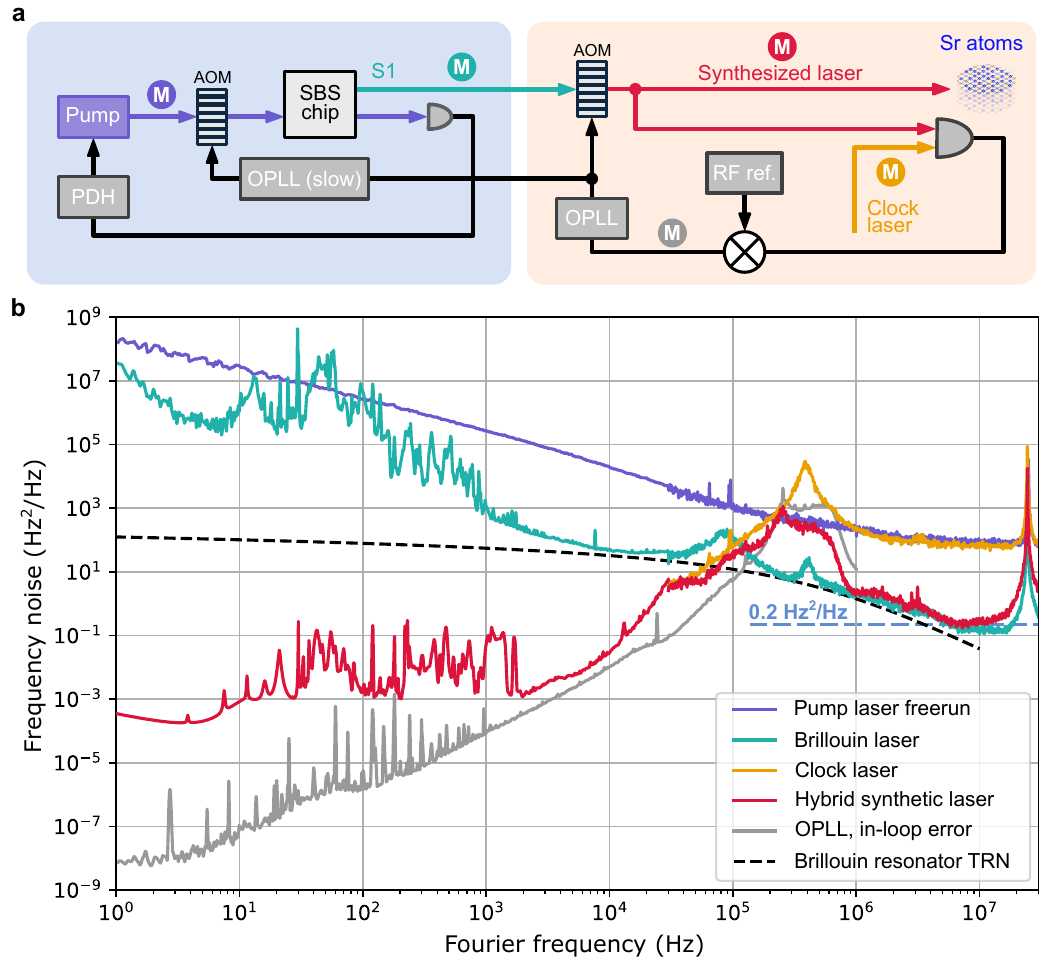}
\caption{\textbf{Frequency noise measurements of the hybrid synthetic laser.} 
\textbf{a} Simplified schematic of the hybrid synthetic laser and the frequency noise measurement locations (circle with ``M''). \textbf{b} Frequency noise spectra of the pump laser (purple), first-order Stokes Brillouin laser (turquoise), the clock laser (orange), the hybrid synthetic laser (red) and the in-loop error signal of the optical phase-lock loop (gray). The frequency noise of the hybrid synthetic laser below 40 kHz is estimated from the clock laser and therefore overlaps with it. The horizontal blue dashed line indicates a frequency noise power spectral density of 0.2 Hz$^2$/Hz. The validity of this estimation is confirmed by the in-loop error signal and Rabi spectroscopy. The TRN floor (dashed black) of the on-chip SBS cavity is included for reference.}\label{fig4}
\end{figure}

The resulting frequency noise is shown in Fig.~\ref{fig4}. The low- and high-Fourier frequency-regions are measured using separate techniques with the resulting noise profiles stitched together at 40 kHz to obtain the full spectrum from 1 Hz to 30 MHz (details in Methods). The frequency noise is measured at different stages of the hybrid synthetic laser, with corresponding measurement locations marked in Fig.~\ref{fig4}a. The high-Fourier-frequency noise is measured using a delayed self-homodyne technique with a fiber Mach–Zehnder interferometer (MZI), whose 24.4 MHz FSR introduces characteristic features in the frequency noise spectrum (Fig.~\ref{fig4}b). The low-Fourier frequency noise of the pump and the Brillouin laser is obtained by measuring the phase noise of a heterodyne beat note between each laser and the stable clock laser~\cite{yan_high-power_2025}. The noise of the hybrid synthetic laser cannot be measured directly because it inherits the stability of the clock laser in this frequency range. Therefore, we confirm the stability transfer from the clock by ensuring that the in-loop noise of the OPLL beat is below the clock laser noise over the entire loop bandwidth. Below 1 kHz, the clock laser noise performance was independently characterized with an atomic spectrum analyzer \cite{yan_high-power_2025} and above 1 kHz, by cross-correlating it with two other ultra-stable lasers~\cite{lee_frequency_2026}.

The frequency-noise spectra in Fig.~\ref{fig4}b illustrate the noise-suppression performance at different stages of the hybrid synthetic laser system. The Brillouin laser reduces the noise from the pump laser (purple) down to 0.2 Hz$^2$/Hz at Fourier frequencies above 7 MHz (turquoise), and to the resonator TRN limit (dashed black) in the 10 kHz--7 MHz range. The pump laser high-Fourier-frequency noise is suppressed by 27 dB, which matches theoretical predictions based on the phonon and resonator decay rates (see Methods). At lower Fourier frequencies, the noise is limited by thermal noise associated with the SBS process and by the stability of the resonator. Therefore, below 100 kHz Fourier frequency, the stability is provided by locking to the clock laser derived from the cryogenic silicon cavity. While the original clock-laser frequency noise (SBS pump phase-locked to the clock-laser; orange) follows the free-running pump-laser noise above the lock bandwidth, the hybrid synthetic laser (red) shows the same level of frequency noise as the Brillouin laser above 1 MHz. Noise between 100 kHz and 1 MHz arises from the OPLL servo bump but remains below that of the pump laser owing to SBS noise suppression. The in-loop error of the OPLL (gray) lies well below the clock-laser frequency noise over the whole bandwidth up to $>200$ kHz, indicating that the loop provides sufficient gain to reduce the Brillouin laser frequency noise to the clock-laser level.

\begin{figure}[h!]
\centering
\includegraphics[width=1.0\textwidth]{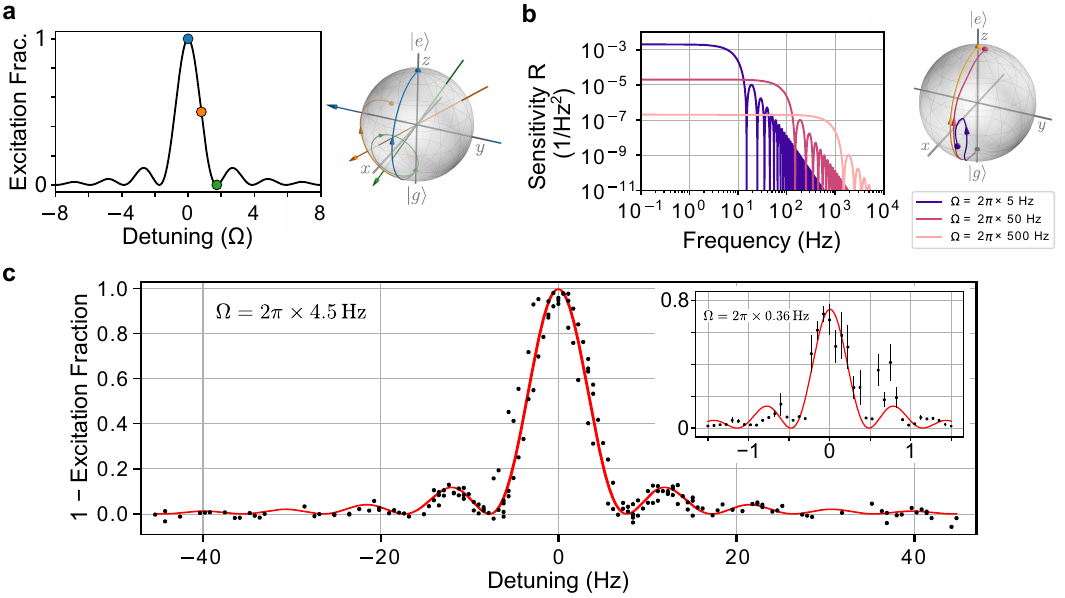}
\caption{\textbf{Rabi spectroscopy on the $^{87}$Sr clock transition.} \textbf{a} Schematic of the excitation fraction for a Fourier-limited Rabi spectroscopy between ground state $\vert g\rangle$ and excited state $\vert e\rangle$ based on a resonant $\pi$-pulse (black line) at bare Rabi frequency $\Omega$ and the corresponding Bloch vector trajectories. In the rotating frame of the laser, increasing detuning tilts the rotation axis (straight arrows) from the $y$-axis towards the $z$-axis and increases the effective Rabi frequency. \textbf{b} Sensitivity functions $R(f)$ of Rabi spectroscopy to constant laser frequency noise PSD of 1 Hz$^2$/Hz at different Rabi frequencies and the corresponding simulated Bloch vector trajectories. This relates the noise PSD $S_\nu(f)$ to the infidelity of the resonant $\pi$-pulse: $1-{\cal F}=4\pi^2\int S_\nu R\,df$ \cite{jiang_sensitivity_2023}. The spectroscopy sequence acts as a low-pass filter with increased sensitivity to laser noise at lower Rabi frequencies. This highlights the importance of suppressing low-Fourier-frequency noise in usual clock Rabi spectroscopy at small $\Omega$, and high-Fourier-frequency noise for more complex pulse sequences and fast qubit operations at large $\Omega$. \textbf{c} Experimentally measured Rabi spectroscopy on $^{87}$Sr trapped in a three-dimensional lattice using the hybrid synthetic laser at Rabi frequency $\Omega=2\pi\times4.5$ Hz (black dots). The atoms are prepared in the excited state $^3$P$_0$, $m_\text{F}=-9/2$ and driven to the ground state $^1$S$_0$, $m_\text{F}=-9/2$. The red line is a fit of the Fourier-limited Rabi line (cf.~panel (a)) with line center detuning and excitation amplitude as free parameters. The inset shows another narrow-line Rabi spectroscopy at $\Omega=2\pi\times0.36$ Hz; we estimate that magnetic field noise induces the excess fluctuations visible in the data (see Methods for details). Error bars indicate the $\pm1\sigma$ interval of the standard error of the mean.}\label{fig5}
\end{figure}

Next, we use this hybrid laser to perform Rabi spectroscopy on a three-dimensional $^{87}$Sr lattice clock. This measurement provides independent confirmation of the low-Fourier-frequency laser performance and demonstrates application of this visible light hybrid laser to quantum sensing and computing applications. The Rabi spectroscopy response scales with the Rabi frequency $\Omega$ when $\Omega$ is greater than the atomic linewidth  (Fig.~\ref{fig5}a). For increasing relative detunings $\delta/\Omega$ from the atomic resonance, the Rabi rotation axis of the Bloch vector tilts towards the south pole, leading to a reduced amplitude of the atom response. The resonant Rabi $\pi$-pulse applied to the atoms becomes increasingly sensitive to low-frequency noise as the Rabi frequency $\Omega$ is reduced. This makes Rabi spectroscopy a particularly sensitive probe for laser noise at frequencies below the Rabi drive frequency while being increasingly insensitive to noise above $\Omega$. Thus, for fast qubit manipulations this highlights the importance of high-frequency laser noise suppression. As an example, Fig.~\ref{fig5}b shows sensitivity spectra for different Rabi frequencies. Simulated Bloch vector traces for a resonant $\pi$-pulse subject to a white frequency noise PSD at 1 Hz$^2$/Hz illustrate the impact of the low-Fourier-frequency noise.

To perform Hz-level Rabi spectroscopy, approximately 40 µW of the hybrid synthetic laser is delivered to the Sr clock via a path-length-stabilized fiber link. This laser power yields a Rabi frequency of $\Omega=2\pi\times4.5$ Hz on the clock transition between electronic states $^1S_0$ and $^3P_0$ with hyperfine spin projection $m_\text{F}=-9/2$. The overlap of the measured excitation fractions with Fourier-limited excitation profile (red line) shown in Fig.~\ref{fig5}c confirms that the hybrid synthetic laser suppresses the low-Fourier-frequency noise as expected from the data presented in Fig.~\ref{fig4}b. Further spectroscopy at a lower Rabi frequency of 0.36 Hz shown in the inset also certifies a sub-Hz laser phase-integrated linewidth (Fig.~\ref{fig5}c inset). The spectroscopy data is recorded over a time frame of more than 2 hours without any feedback on the laser frequency.

\section{Summary of Results and Outlook}\label{sec3}

We demonstrate an ultra-low-noise hybrid synthetic laser that achieves a sub-Hz phase-integrated linewidth and a frequency noise of 0.2 Hz$^2$/Hz at high-Fourier-frequencies. The hybrid laser combines a visible light 698 nm on-chip Brillouin laser with a cryogenic Si reference cavity to achieve ultra-low-noise laser performance over 7 decades of Fourier frequency. We further confirm the sub-Hz integral linewidth and demonstrate application of this laser to a quantum sensing application by performing Rabi spectroscopy on the three-dimensional $^\text{87}$Sr lattice clock transition. By introducing a small form factor visible light photonic chip as a high-Fourier-frequency noise suppressor without degrading the performance of the Si-cavity-stabilized laser, this approach extends the frequency noise suppression bandwidth while preserving the ultra-low-noise characteristics of state-of-the-art reference cavities. The hybrid design provides a broadband frequency noise engineering approach to precision laser design~\cite{yan_high-power_2025}. 

The integrated visible light (698 nm) coil-Brillouin laser enables direct interfacing with the cryogenic Si reference cavity and the atom lattice without intermediate wavelength conversion and introduction of further noise sources. The thin core Si$_3$N$_4$ platform provides multiple important features that enable lasing in the visible and large-mode volume coil-Brillouin laser design to suppress the noise and increase the optical output power. The low frequency noise benefits from the low TRN floor of the 65-cm-long Brillouin coil resonator cavity in the $\sim$10 kHz to 1 MHz Fourier frequency range compared to previous shorter Brillouin resonators in the visible range \cite{chauhan_visible_2021,song_octave_2026}. Additionally, the Brillouin laser chip is self-isolating, simplifying the experimental setup and facilitating future integration with other components in the Si$_3$N$_4$ platform. For example, integration of the clock laser and control modulators \cite{wang_silicon_2022} directly on the photonic chip can further improve reliability and robustness as well as lower weight and cost. 

Future improvements of the hybrid synthetic laser output power are essential for increasing the Rabi frequency to provide benefits from noise reduction at high Fourier-frequencies for narrow clock transitions. Higher output power and better noise suppression of photonic-chip-based Brillouin laser may be enabled by longer coil resonators and more efficient fiber-chip coupling designs. An alternative and complementary approach is to implement S2 suppression \cite{liu_integrated_2024,wang_taming_2024}, which enables higher S1 power without incurring a threshold penalty. Additional power scaling while preserving the Brillouin laser noise performance can be achieved using coherent amplification \cite{kuznetsov_ultrabroadband_2025,zhao_ultrabroadband_2025}, coherent beam combining approaches \cite{nagano_multistage_2024,zeng_wattlevel_2022}, and injection locking to a high power laser diode \cite{kim_optical_2015}. Another approach is to employ a Brillouin amplifier using an additional on-chip Brillouin resonator pumped just below threshold \cite{otterstrom_resonantly_2019}. For example, a Brillouin resonator with longer length could be a candidate due to increased threshold and sufficiently good amplifier pump noise suppression to avoid added amplifier noise \cite{debut_linewidth_2000}. In parallel, photonic chip heterogeneous packaging with reliable high optical power fiber coupling at visible wavelengths are complementary avenues towards optimized efficiency and performance. 

These results point the way to new laser frequency noise engineering approaches that combine chip-scale integration with precision bulk-scale technologies, enabling a new regime of quantum sensing and computation where the inherent stability of today's systems is augmented with fast, coherent quantum state control. Further integration of these hybrid lasers can enable transfer of this technology to portable and low cost quantum clock and sensing applications.

\section{Methods}\label{sec4}

\subsection{Brillouin laser design, fabrication, and setup}
The on-chip Brillouin resonator device employs the TM$_0$ mode in a 40 nm thick ultra-low-loss Si$_3$N$_4$ waveguide with a 15 µm lower and a 6 µm upper SiO$_2$ cladding. The waveguide is 2.3 µm for the coil Brillouin resonator and 3 µm for the filter ring. The loaded quality factor (Q) of the Brillouin resonator is 38 million and the total linewidth of the resonance is 11 MHz. The loaded Q of the filter resonator is 2.8 million and the total linewidth of the resonance is 154 MHz. The filter ring linewidth is narrower than the Brillouin resonator FSR (291 MHz) to ensure overlap with only a single Brillouin resonance, while remaining wide enough to allow easy alignment.

The photonic integrated chip is fabricated on silicon substrates with a 15 µm thermally grown silicon dioxide lower cladding. A stoichiometric Si$_3$N$_4$ layer is deposited by low-pressure chemical vapor deposition (LPCVD), followed by waveguide patterning using a deep ultraviolet (DUV) stepper and transfer by inductively coupled plasma reactive ion etching (ICP-RIE). The upper cladding consists of a 6 µm silicon dioxide layer deposited by tetraethylorthosilicate (TEOS)-based plasma enhanced chemical vapor deposition (PECVD) and subsequently annealed to reduce optical scattering and absorption losses.

 The Brillouin lasing light is generated by locking the 698 nm pump laser to the 65-cm integrated cavity (see Fig.~\ref{fig2}a) via the Pound-Drever-Hall (PDH) scheme \cite{drever_laser_1983}. The pump light transmitted by the Brillouin resonator provides the error signal. The pump laser power is kept under S2 threshold to prevent added noise from cascaded Brillouin lasing. All chip waveguide I/O interfaces are routed to a single location at the chip edge and coupled with a single polarization-maintaining fiber array (Fig.~\ref{fig3}c). The photonic chip is mounted on a temperature-stabilized stage with 1 mK precision and enclosed in a compact metal housing to minimize air-flow-induced disturbances.  

\subsection{Frequency noise measurement}
High Fourier-frequency-noise is measured using a delayed self-homodyne technique with a fiber Mach–Zehnder interferometer (MZI). The unbalanced interferometer is first calibrated to determine its FSR. The laser under test is then injected into the interferometer, and the differential optical power at the two outputs is detected with balanced photodetectors. By ramping the phase in one interferometer arm, the peak-to-peak voltage response of the photodetector output is measured. Together with the calibrated FSR, this voltage swing provides a conversion factor between detected power fluctuations and optical frequency deviations near the quadrature point. During noise measurements, the system is triggered at the quadrature point and sampled at different rates to cover distinct Fourier frequency ranges. Multiple acquisitions are averaged, and data obtained at different sampling rates are stitched together to obtain the full frequency noise spectrum. At low-Fourier frequencies, the measurement sensitivity is limited by thermal and acoustic noise in the optical fibers. As a result, the frequency noise spectrum below 30 kHz is truncated. For frequency noise below 30 kHz, the clock laser is used as a reference when applicable. The phase noise of the optical beat note between the test laser and the clock laser is measured using a phase noise analyzer and subsequently converted to frequency noise.

\subsection{Brillouin laser pump noise suppression ratio}
The pump noise suppression ratio can be predicted by the equation below \cite{dallyn_thermal_2022} based on the phonon decay rate and the Brillouin resonator decay rate. 
\begin{equation}
    S_f^{\text{trans}}[\omega] \approx \frac{\gamma^2}{\Gamma^2} S_f^{\text{pump}}[\omega]
\end{equation}
$S_f^{\text{trans}}[\omega]$ is the pump transfer noise and $S_f^{\text{pump}}[\omega]$ is the pump laser frequency noise. $\gamma$ and $\Gamma$ are the decay rates of the resonator and the phonons. This calculation is under the strongest suppression condition of pumping just below S2 threshold. In our device, the resonator decay rate, or the linewidth of the resonance, is 11 MHz. The phonon decay rate, estimated from the Brillouin gain bandwidth of this waveguide at 698 nm, is 300 MHz. These parameters yield a pump laser noise suppression ratio of 28.7 dB. The demonstrated 27.0 dB suppression is very close the theoretical prediction. We expect the small difference to originate from inaccuracies in the phonon decay rate, resonance linewidth variation, and the pump laser power not reaching just below S2 threshold.

\subsection{Fluctuations in Rabi spectroscopy}
Independent Rabi spectroscopy with the same clock laser at similar Rabi frequencies on a different system utilizing a one-dimensional $^{87}$Sr lattice clock confirmed an absence of excess fluctuations as observed in the inset of Fig.~\ref{fig5}c. Additionally, performing Rabi spectroscopy on the three-dimensional clock system with the clock laser showed a similar level of fluctuations. This demonstrates that the source of the fluctuations do not originate from the laser, but instead stem from the atomic system itself. Further investigation by performing spectroscopy on differently magnetically sensitive transitions between different hyperfine levels revealed that the excess fluctuations were induced by mG-level noise on the 2.92 G magnetic offset field between experimental realizations. On the $m_\text{F}=-9/2\rightarrow m_\text{F}=-9/2$ transition employed in the experiment with a Zeeman splitting of approx.~$-670$ Hz/G this explains the observed scatter in the excitation fraction. Apart from fluctuations in the offset magnetic field, these variations might also originate from percent-level variations in the circularity of one of the lattice beams, inducing a vector Stark shift of a corresponding amplitude on the Sr clock transition.

\bibliography{sn-bibliography}

\backmatter

\section*{Acknowledgments}
This work is supported by funding from the the NSF under QLCI Q-SEnSE award number 2016244 and the Army Research Office (ARO) AMP program under award W911NF2310179. The views and conclusions contained in this document are those of the author(s) and should not be interpreted as representing the official policies of DARPA or the U.S.~government. The authors thanks Karl D.~Nelson of Honeywell Aerospace for help in the device fabrication and S.L.~acknowledges funding from the Alexander von Humboldt Foundation.

\section*{Declarations}

\begin{itemize}

\item Conflict of interest/Competing interests

Dr. Blumenthal has consulted for Infleqtion and owns stock in the company.  The other authors declare no potential conflict of interest.

\item Data availability 

The data that support the plots within this paper and other findings of this study are available from the corresponding author on reasonable request.

\item Supplementary Information

Supplementary Information is not available.

\item Author contribution

 M.S.~and N.C.~designed the integrated photonic chips. M.S., A.I., and N.M.~characterized the photonic device. M.S., D.L., L.Y., A.I., S.L., and N.M.~built the hybrid synthetic laser and performed laser frequency noise experiments. S.L., L.Y., M.N.F, Y.H.L, S.A., and M.S.~performed the Rabi spectroscopy measurements. All authors contributed to analyzing the experimental results. M.S.~and S.L.~prepared the manuscript. J.Y.~and D.J.B.~supervised and led the scientific collaboration.
\end{itemize}

\end{document}